\documentclass[aps, pre, eps, superscriptaddress, twocolumn, amsmath,amssymb,floatfix]{revtex4}

\newif\ifpdf\ifx\pdfoutput\undefined\pdffalse\else\pdfoutput=1\pdftrue\fi

\include{eps.cls}

\usepackage{graphicx}
\usepackage{subfigure}
\usepackage{epsfig}
\begin{document}

\title{\bf Polydisperse hard spheres at a hard wall}

\author{Matteo Buzzacchi}
\affiliation{Department of Physics, University of Bath, Bath BA2 7AY,
United Kingdom}

\author{Ignacio Pagonabarraga}
\affiliation{Universitat de Barcelona, Dept F\`{i}sica Fonamental, E-08028 Barcelona, Spain}

\author{Nigel B. Wilding}
\affiliation{Department of Physics, University of Bath, Bath BA2 7AY, United Kingdom}

\date{\today}

\begin{abstract} 

The structural properties of polydisperse hard spheres in the presence
of a hard wall are investigated via Monte Carlo simulation and density
functional theory (DFT). Attention is focussed on the local density
distribution $\rho(\sigma,z)$, measuring the number density of
particles of diameter $\sigma$ at a distance $z$ from the wall. The
form of $\rho(\sigma,z)$ is obtained for bulk volume fractions
$\eta_b=0.2$ and $\eta_b=0.4$ for two choices of the bulk parent
distribution: a top-hat form, which we study for degrees of polydispersity
$\delta=11.5\%$ and $\delta=40.4\%$, and a truncated Schulz form having
$\delta=40.7\%$. Excellent overall agreement is found between the DFT
and simulation results, particularly at $\eta_b=0.2$. A detailed
analysis of $\rho(\sigma,z)$ confirms the presence of oscillatory size
segregation effects observed in a previous DFT study (Pagonabarraga
{\em et al.}, Phys. Rev. Lett. {\bf 84}, 911 (2000)). For large
$\delta$, the character of these oscillation is observed to depend
strongly on the shape of the parent distribution. In the vicinity of
the wall, attractive $\sigma$-dependent depletion interactions are
found to greatly enhance the density of the largest particles. The
local degree of polydispersity $\delta(z)$ is suppressed in this region,
while further from the wall it exhibits oscillations.
 
%Compared to the monodisperse limit, the improved overall
%packing of polydisperse fluids is found to dampen wall-induced density
%oscillations.

\end{abstract} 
\maketitle
\setcounter{totalnumber}{10}
\section{Introduction and background}
\label{sec:intro}

Many complex fluids, whether natural or synthetic in origin, are
intrinsically polydisperse in character, that is they comprise mixtures
of similar rather than strictly identical constituent particles. Common
examples are colloidal dispersions, macromolecules in solution and
liquid crystals, where variation amongst the particles can occur in
attributes such as their size, shape or surface charge. 
Polydispersity is of considerable practical relevance because it can
affect the properties of materials in a variety of applications such 
as coating technologies \cite{COATINGS}, foodstuffs \cite{LARSON99} and
polymer processing \cite{PROCESSING}. However, from the fundamental
perspective, the current understanding of polydisperse fluids is considerably
less advanced than that of their monodisperse counterparts. The reason
for this is the inherent complexity that polydispersity imparts to a
system, and which generates difficulties for experimental, theoretical and
simulation approaches alike.

On the experimental front, intricate technical issues beset the
preparation, characterization and analysis of polydisperse samples, and
only relatively recently has work begun to systematically address the
generic consequences of polydispersity, such as its effects on phase
behaviour and the fractionation of particles of different sizes between
coexisting phases \cite{FAIRHURST04,Experimental,PUSEY}. As far as
analytical theory is concerned, the principal challenge is the
multitude of variables required to accurately describe the system's
properties. Statistical mechanical theories of polydispersity typically
regard the polydisperse attribute as a continuous variable ($\sigma$,
say) \cite{GUALTIERI82,SALACUSE}. Accordingly the system may be
regarded as a mixture of an infinite number of species--each labelled
by the value of $\sigma$.  It is then natural to describe the
polydispersity in terms of a density distribution, $\rho^0(\sigma)$,
measuring the number density of each species. Difficulties  arise,
however, when one attempts to determine the thermodynamic properties of
the system, such as its phase behaviour. Because the free energy
depends on (is a functional of) the entire distribution
$\rho^0(\sigma)$, it occupies a parameter space that is effectively
infinite dimensional. This in turn renders the analysis of phase
behaviour much more problematic than for monodisperse systems or finite
mixtures of a few components. 

For small degrees of polydispersity, perturbative approaches permit
some theoretical headway to be made \cite{RMLE99,RMLE01}. A more
general approach is the moment free energy (MFE) method
\cite{SOLLICH98,WARREN98,SOLLICH01,SOLLICH02}.  For certain truncatable
forms of the free energy use of this method allows the effective
dimensionality of the problem to be reduced to a manageable level by
projecting the free energy onto a suitably defined subspace spanned by
a few principal moments of $\rho^0(\sigma)$. The phase behaviour of the
resulting projected free energy can be calculated relatively
straightforwardly. This method has been applied to investigate a range
of open questions related to bulk phase equilibria, including {\em
(inter-alia)} issues concerned with liquid-vapor coexistence and critical
point shifts \cite{WILDING04}, freezing \cite{FASOLO03,FASOLO04} and
liquid crystal phase transitions \cite{SpeSol02}.

These advances in theoretical methods for dealing with polydisperse
fluids, have recently been complemented by parallel developments in
simulation methodologies. New Monte Carlo (MC) algorithms now permit
the effective study of polydisperse phase equilibria within the grand
canonical ensemble (GCE). Use of this ensemble is advantageous for the
study of phase transitions because it allows the system density to
fluctuate as a whole, thereby catering naturally for both order
parameter fluctuations and fractionation effects \cite{WILDING02}.
Within the GCE framework, polydispersity is controlled by means of a
chemical potential distribution $\mu(\sigma)$, the form of which is
adapted (for each state point of interest) in such a way as to yield an
ensemble averaged density distribution $\bar\rho(\sigma)$ that matches
some prescribed form $\rho^0(\sigma)$. This approach extends previous
ones (see eg. refs.\cite{KOFKE87,BATES98}) by facilitating the
targeting of a {\em specific} density distribution. It thus corresponds
more closely to the experimental situation for polydisperse fluids such
as colloids and polymers, where the form of $\rho(\sigma)$ is {\em
fixed} by the synthesis of the fluid and only its scale can vary
depending on the quantity of solvent present. The new techniques have
recently been applied to obtain the equation of state of polydisperse
hard spheres \cite{WILDING02} and to investigate polydispersity-induced
alterations to bulk phase behaviour in a Lennard-Jones fluid
\cite{WILDING04}. 

While there is considerable ongoing progress in the study of bulk phase
behaviour, to date comparatively little work has been reported
regarding the influence of polydispersity on the properties of
inhomogeneous fluids \cite{BARRAT}. One of the standard contemporary
theoretical approaches for dealing with such systems is density
functional theory (DFT) \cite{EVANS92}. Recently, it has been shown by
one of us, that the MFE method outlined above carries over
straightforwardly to certain well established density functionals
\cite{PAGON00,PAGON01}. The first such study of an inhomogeneous system
considered the prototype model for a inhomogeneous polydisperse fluid,
namely hard spheres  in the vicinity of a hard wall \cite{PAGON00}.
Novel oscillatory behaviour was reported in the local concentration of
different sized species as a function of the wall distance. Other DFT
work has investigated polydisperse polymer/solvent mixtures at
coexistence \cite{PAGON01}, and the effect of polydispersity on the
fractionation and surface tension of a liquid-gas interface within a
van der Waals approximation \cite{BELLIER02}.

The extension of DFT techniques to polydisperse fluids is a welcome
development. However, as yet, the reliability of the approximations
inherent in such treatments remain untested. Clearly, therefore, it is
desirable to obtain benchmarks with which these and other theories of
confined polydisperse fluids can be compared. In this paper we address
this issue using MC simulation and DFT applied to size-disperse hard
spheres at a hard wall. For this purpose the GCE simulation methods
developed for the study of the bulk phase equilibria of polydisperse
fluids carry over directly, allowing us to probe confinement-induced
changes to the density distribution. In addition to facilitating a
detailed comparison between simulation and DFT, our results confirm the
presence of the oscillatory size segregation effects seen previously in
the original DFT study, as well as further elucidating the character of
the local fluid structure and the effects of depletion interactions near
the wall.

%The phase and structural behaviour of binary and ternary hard sphere
%mixtures, in bulk or confined geometries has also been analyzed by
%theory~\cite{DIETRICH00},\cite{PATRA99} and experiments,  showing some
%interesting phenomena, like entropy-driven fluid-solid transitions in 
%a dispersion of polystyrene spheres at a glass
%wall~\cite{Experimental},  and have helped clarify the nature of
%entropic forces in a system of particles interacting through a purely
%repulsive potential.

Our article is organized as follows: in Sec.~\ref{sec:method} we
outline the MC simulation scheme and the density functional theory. In
Sec.~\ref{sec:results} we analyse and compare in detail the MC and DFT
results for two choices of the parent size distribution at two volume
fractions.  Finally, a discussion of the results and the prospects for
interesting further work features in Sec.~\ref{sec:concs}.

\section{System and Methodology}
\label{sec:method}

The system with which we shall be concerned is a fluid of hard spheres,
whose polydisperse attribute $\sigma$ corresponds to the sphere
diameter. Pairs of particles $i$ and $j$ are assumed to interact via
the potential

\begin{equation}
U_{pp}({\bf r}_{i},{\bf r}_{j},\sigma_{i},\sigma_{j})=\left\{
\begin{array}{ll}
+\infty & \mbox{ if $|{\bf r}_{i}-{\bf r}_{j}|<\frac{\sigma_{i}+\sigma_{j}}{2}$} \\
~~0      & \mbox{ otherwise }
\end{array}
\right..
\end{equation}

Consider initially such a system in the bulk. Conventionally
the state of the system is described by a ``parent'' density
distribution \cite{SOLLICH02,NOTE0} $\rho^0(\sigma)$, which can be written

\begin{equation}
\rho^0(\sigma)=n_0f(\sigma)\:.
\label{eq:parent}
\end{equation}
Here $n_0=N/V$ is the overall particle number density, while $V$ is the
system volume and $f(\sigma)$ is a normalized shape function, the
relative width of which is quantified by the
dimensionless degree of polydispersity:

\begin{equation}
\delta=\frac{\sqrt{\overline{(\sigma-\bar{\sigma})^2}}}{\bar\sigma} \:,
\label{eq:poly}
\end{equation}
measuring the standard deviation of the parent distribution, normalized by its mean.

Let us now break the translational symmetry of the system in one
coordinate direction ($z$, say) by introducing a hard wall in the plane
$z=0$. Interactions of the particles with the wall are assumed to be
controlled by the potential:

\begin{equation}
U_{pw}(z_i,\sigma_i)=\left\{
\begin{array}{ll}
+\infty & \mbox{ if $z_{i}<\sigma_{i}/2$} \\
~~0      & \mbox{ otherwise }
\end{array}
\label{eq:upw}
\right..
\end{equation}
Near such a wall, modifications to bulk behaviour arise even in
monodisperse systems from standard packing effects. For polydisperse
systems, however, additional factors come into play. Firstly, the
excluded volume constraint associated with $U_{pw}$ implies that the
centers of small particles are permitted to approach the wall more
closely than those of large particles. Hence one expects (for
sufficiently small $z$) that the local density distribution will be
truncated at large $\sigma$ with respect to the bulk (parent) form.
Secondly, experience with binary hard sphere mixtures \cite{EVANS01}
suggest that very close to the wall one can expect attractive depletion
interactions between the particles and the wall. In the polydisperse
case these depletion forces should be $\sigma$-dependent. In order to
quantify the net influence of all these effects,  we consider the
ensemble averaged local density distribution $\rho(\sigma,z)$ at a
perpendicular distance $z$ from the wall:

\begin{equation}
\rho(\sigma,z)=\int_0^{L_x}  \int_0^{L_y} \rho(\sigma,{\bf r})\; dx dy\:.
\end{equation}

\begin{figure}[h]
\includegraphics[width=7.0cm,clip=true]{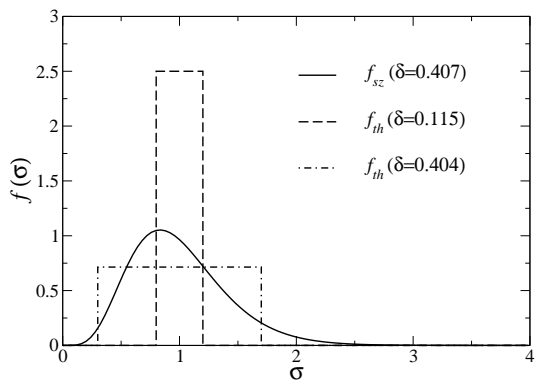}
\caption{The top-hat and Schulz distributions considered in this work.}
\label{fig:dists}
\end{figure}

In the present work, we have elected to study two forms of the parent
distribution: a top-hat distribution and a Schulz, the form of which
are shown in fig.~\ref{fig:dists}.  The top-hat distribution  is
defined by 

\begin{equation}
f_{th}(\sigma)=\left\{
\begin{array}{ll}
(2c)^{-1} & \mbox { if $1-c\le \sigma \le 1+c$} \\
~~0      &  \mbox { otherwise }
\end{array}
\right.,
\end{equation}
where (without loss of generality), the mean particle diameter has been
set to $\bar{\sigma}=1.0$. We have studied the cases $c=0.2$ and $c=0.7$, for
which $\delta=c/\sqrt{3}\approx 0.115$ and $\delta\approx 0.404$ respectively. 

The Schulz distribution is defined by the shape function

\begin{equation}
f_{sz}(\sigma)=\frac{1}{z!}\left(\frac{z+1}{\bar{\sigma}}\right)^{z+1}\sigma^z\exp\left[-\left(\frac{z+1}{\bar{\sigma}}\right)\sigma\right]\:.
\end{equation}
Here the parameter $z$ controls the width of the distribution and
thence the value of $\delta$. We have considered the case $z=5$,
corresponding to $\delta=(z+1)^{-1/2}\simeq 0.408$. Again, by
construction, $\bar{\sigma}=1.0$. Notice that in contrast to eg. a
Gaussian, the Schulz distribution vanishes smoothly (has a natural
cutoff) as $\sigma\to 0$. For the purposes of the MC simulations
described below, however, one does require an upper cutoff in $\sigma$.
We chose to truncate $f(\sigma)$ at $\sigma_c=4.0$,  which reduces the
degree of polydispersity to $\delta\approx 0.407$. The truncated form
of $f(\sigma)$ was then rescaled to unit integrated area. Given this
truncation, the population of particles at the cutoff diameter is
very small compared to those having the mean diameter:
$\rho^0(\sigma_{c})/\rho^0(\bar{\sigma})=1.7\times 10^{-5}$. However,
the volume of the largest allowed particle is $64$ times that of one
having the mean diameter. 

A quantity useful in characterizing a polydisperse system is its volume
fraction, which provides a measure of the degree to which the particles
pack the available space. We shall employ it in preference to the
number density $n_0$ which does not alone provide equivalent
information. We define the volume fraction in terms of the {\em bulk}
parent distribution:

\begin{equation}
\eta_b=\frac{\pi}{6}\int~d\sigma~\rho^0(\sigma)\sigma^3.
\end{equation}
For a nominated shape function $f(\sigma)$, choosing a value for
$\eta_b$ (or equivalently $n_0$) serves to fix the bulk chemical
potential distribution $\mu(\sigma)$. The assumption that the confined
fluid exists in equilibrium with a bulk reservoir then implies that
both the bulk and confined systems have equal $\mu(\sigma)$. Note
however, that for the confined hard sphere systems we consider here,
the system-averaged volume fraction is generally less than that of the
reservoir.  For both the top-hat and Schulz parents described above, we
consider the cases $\eta_b=0.2$ and $\eta_b=0.4$. 

With regard to our use of scales, we shall throughout express all
lengths in units of the average particle diameter $\bar\sigma$ of the
parent distribution. Owing to the lack of a temperature scale in hard
sphere systems, we implicitly adopt the convention of assigning
$\beta=1/k_BT\equiv 1$. 

\subsection{Grand-canonical ensemble Monte Carlo calculations}

The Grand Canonical Metropolis Monte Carlo algorithm we have employed
has previously been described in ref.~\cite{WILDING02} and involves
particle insertions, deletions, translations, and resizing
operations.  Particle resizing operations $(\sigma\rightarrow\sigma')$,
or the insertion of a new particle are performed by drawing a random
number with uniform probability in the range of allowed sizes.
Acceptance/rejection criteria for these moves involve not only the
change in the particle-particle and particle-wall potentials, but also
the change in the chemical potentials $\mu(\sigma)$ and $\mu(\sigma')$.
 Translational moves $(\sigma,{\bf r})\rightarrow(\sigma,{\bf r}')$ 
are not strictly necessary in the GCE formalism, as they are equivalent
to the successive  deletion of particle $(\sigma,{\bf r})$ and
insertion of particle $(\sigma,{\bf r}')$, but their use proved
beneficial to the sampling efficiency at the higher local volume
fractions encountered near the wall.

Within the GCE framework, the form of the ensemble averaged density
distribution $\bar\rho(\sigma)$ is controlled by its conjugate chemical
potential distribution $\mu(\sigma)$. In order to perform simulations
of a prescribed parent distribution one therefore needs to match
$\bar\rho(\sigma)$ to  $\rho^0(\sigma)$ in a bulk (or, in our case,
fully periodic) simulation. Unfortunately, the task of determining the
requisite $\mu(\sigma)$ is complicated by the fact that it is a {\em
functional} of $\rho^0(\sigma)$. Effectively therefore, one is faced
with solving the inverse problem $\mu(\sigma)=\mu[\rho^0(\sigma)]$.
Doing so is facilitated by a recently proposed MC scheme--the
non-equilibrium potential refinement (NEPR) algorithm
\cite{WILDING03}-- use of which enables the efficient iterative
determination of $\mu[\rho^\circ(\sigma)]$, from a single simulation
run, and without the need for an initial guess of its form. To achieve
this, the method continually updates $\mu(\sigma)$ in such a way as to
correct for the deviation of the instantaneous density distribution
$\rho(\sigma)$ from the target form (i.e.\ the parent). However, tuning
$\mu(\sigma)$ in this manner clearly violates detailed balance. To
counter this, successive iterations reduce the degree of modification
applied to $\mu(\sigma)$, thereby driving the system towards
equilibrium and ultimately yielding the equilibrium form of
$\mu[\rho^\circ(\sigma)]$.

In order to obtain the bulk form of $\mu(\sigma)$ corresponding to the
top-hat and Schulz distributions at $\eta_{b}=0.2$ and $\eta_{b}=0.4$,
the NEPR algorithm was deployed in simulations of a fully periodic
cubic simulation cell of side $L=12$.  The resulting forms of
$\mu(\sigma)$ for the respective parents were then used to study the
effects of introducing two oppositely facing hard walls at $z=0$ and
$z=L$ (the system remaining periodic in the $x$ and $y$ directions).
The cell size $L=12$ proved sufficient to ensure that the fluid
properties at the cell midpoint were indistinguishable (to within
statistical uncertainty) from those of the fully periodic system. This
finding was further tested by comparing the results with those for a
cell that was elongated along the $z$ direction, having $(L_{z}=24)$.
Again within statistical uncertainties, no differences were observed, leading
us to conclude that the two walls do not interact with one another and
that the properties near one wall are representative of the
semi-infinite geometry.

The particle density $\rho(\sigma,z)$ was accumulated in the form of a
histogram in the course of the simulations and will be the object of
our analysis in the next section. The histogram was formed by
discretising the permitted ranges $0\le \sigma \le \sigma_c$ and
$0<z<L_z$, into bins. The bin width in $\sigma$ used for the top-hat
distributions was $\delta\sigma=0.01$, while $\delta\sigma=0.04$ was
employed for the Schulz. The bin width in the $z$ direction was $\delta
z=0.005$ for the top-hat distributions and $\delta z=0.02$ for the
Schulz distribution.

\subsection{Density Functional Theory model}
\label{sec:dft}

The density functional model employed is that introduced by 
Rosenfeld~\cite{ROSENFELD,EVANS92}, following the implementation
proposed in ref.~\cite{ROSINBERG},  and corresponds to a weighted
functional that has proved accurate for monodisperse 
systems~\cite{TARAZONA}. Although improvements have been proposed to
the original Rosenfeld model which appear to perform better in the
context of describing crystalline phases, we do not expect our
implementation to suffer major drawbacks for the regimes of volume
fraction regimes of interest in the present work. 

The free energy functional can be expressed as:

\begin{eqnarray}
\Omega &=& \int d{\bf r} d \sigma \left\{
\rho(\sigma,{\bf r}) \left[
\ln \left(
\Lambda^3(\sigma)\rho(\sigma,{\bf r})\right)-1\right]\right.\nonumber\\
&+&\left.[V(\sigma,{\bf r})-\mu(\sigma)]\rho(\sigma,{\bf r})\right\}+\int d{\bf r}{\cal F}^{ex}[m_{\alpha}({\bf r})]
\label{eq:modelF}
\end{eqnarray}
where the excess contribution to the free energy functional, 
${\cal F}^{ex}=-m_0\ln(1-m_3)+m_1m_2/(1-m_3)+m_2^3/(1-m_3)^2$, is a function of four moments only:

\begin{equation}
m_{\alpha}({\bf r}) = \int d\sigma d{\bf r}'\rho(\sigma,{\bf r}) \omega_{\alpha}(\sigma,|{\bf r}-{\bf r}'|).
\label{eq:moments}
\end{equation}
The four weight functions, $\omega_{\alpha}$ (non-local  in space) are
chosen to recover the Percus-Yevick free energy and correlation
functions for a  homogeneous mixture. In the bulk  the moments have a
simple physical interpretation. Indeed, $m_3(\infty)=\eta_b$,  $m_0(\infty)=\rho_b$,
while  $m_2(\infty)/\rho_b$ is the mean area and $m_1(\infty)/\rho_b$ the mean radius.

The equilibrium density profiles are obtained by  minimizing
the free energy functional, eq.(\ref{eq:modelF}), with respect to the
density profiles, leading to

\begin{equation}
\rho_{eq}(\sigma,{\bf r})=\rho^0(\sigma)\exp\left[-\beta V(\sigma,{\bf r})+\mu^{ex}(\sigma,{\bf r})-\mu^{ex}_b(\sigma)\right]
\label{eq:rhoeq}
\end{equation}
where  $V(\sigma,{\bf r})$ accounts for an external potential, and
$\mu^{ex}_b(\sigma)$ and $\rho^0(\sigma)$ correspond to the excess
chemical potential and density of species $\sigma$ in the
reservoir, respectively.

The advantage of the moment structure of the functional is that the
excess chemical potential of each species is 

\begin{eqnarray}
\mu^{ex}(\sigma,{\bf r})&=&\sum_{\beta} \int d{\bf r}
\omega_{\beta}(\sigma,|{\bf r}-{\bf r}'|)\frac{\partial {\cal
F}^{ex}}{\partial m_{\beta}}({\bf r}') \nonumber\\ &\equiv&
\sum_{\beta} \int d{\bf r} \omega_{\beta}(\sigma,|{\bf r}-{\bf r}'|)
\mu^{ex}_{\beta}({\bf r}')\:,
\end{eqnarray}
and can be interpreted as a linear combination of the four moment
excess chemical potentials, $\mu^{ex}_{\alpha}=\partial {\cal
F}^{ex}/\partial m_{\alpha}$. This moment structure simplifies the
study of polydisperse mixtures \cite{SOLLICH02}. 

For the case of a fluid mixture in the presence of a planar hard
wall, the external potential reduces to eq.(\ref{eq:upw}):
$V(\sigma,{\bf r})=U_{pw}(z,\sigma)$.  The equilibrium density
profiles, as given implicitly  in eq.~(\ref{eq:rhoeq}), are found
numerically via iteration. To this end, it is necessary to define an
underlying lattice and to represent the parent distribution as a set of
discrete species. However, the existence of this lattice imposes a minimum
resolvable length scale (lattice cutoff). To avoid lattice artifacts, one must therefore
apply a lower cutoff to the parent distribution $\rho^0(\sigma)$ which is itself
large compared to the lattice spacing; we take the lower $\sigma$ cutoff
to vary between 20 and 50 lattice units.  Although in principle, this
cutoff may lead to discrepancies between the DFT and MC results (in the
simulations particles can have a vanishingly small radius), we have
verified that the lower cutoff is sufficiently small that changes to its
value have negligible effect on the overall results.

Depending on the degree of polydispersity, the continuous parent
distribution is represented by a few hundred species, ensuring always
that the maximum sized species has a sufficiently small contribution to
avoid spurious effects from the upper $\sigma$ cutoff. For consistency,
we choose the upper cutoff to equal that used in the  simulations. The
mean particle diameter is always taken as the reference length scale in
the calculations. 

\section{Results}
\label{sec:results}

Below we present our findings for the effect of the hard wall on the
density distribution. For the purposes of gauging the accuracy of the
DFT predictions, we have confined our comparison to the
(representative) choice of the narrow top-hat distribution ($c=0.2$,
$\delta=0.115$) and the Schulz distribution ($z=5, \delta=0.407$).
Results for the wide top-hat ($c=0.7 $) have been obtained only via
simulation, but serve to elucidate the effect of distribution shape
changes at a given $\delta$.

\begin{figure}[h]
\centerline{\psfig{file=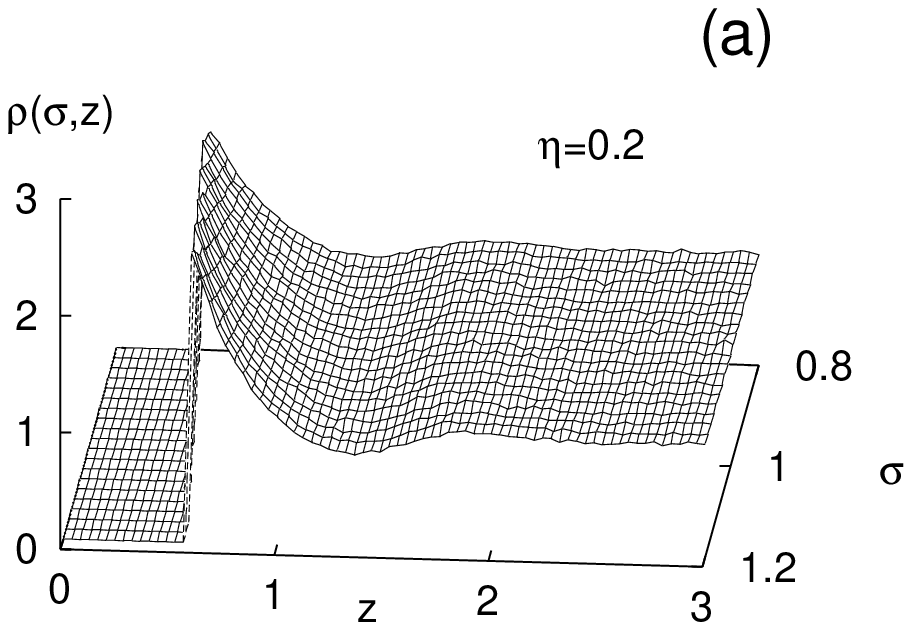,height=5.5cm,width=5.00cm}
\psfig{file=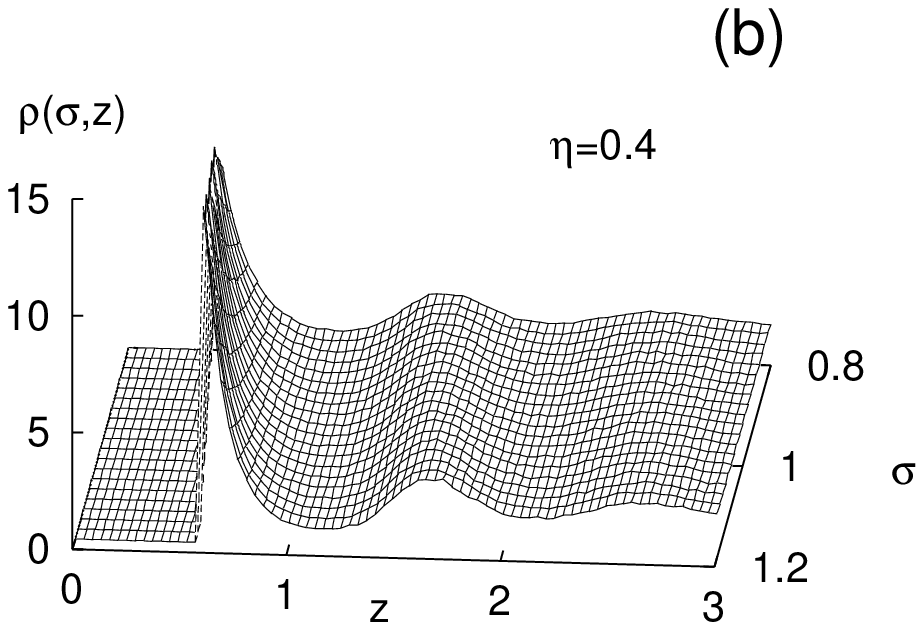,height=5.25cm,width=4.75cm}}
\centerline{\psfig{file=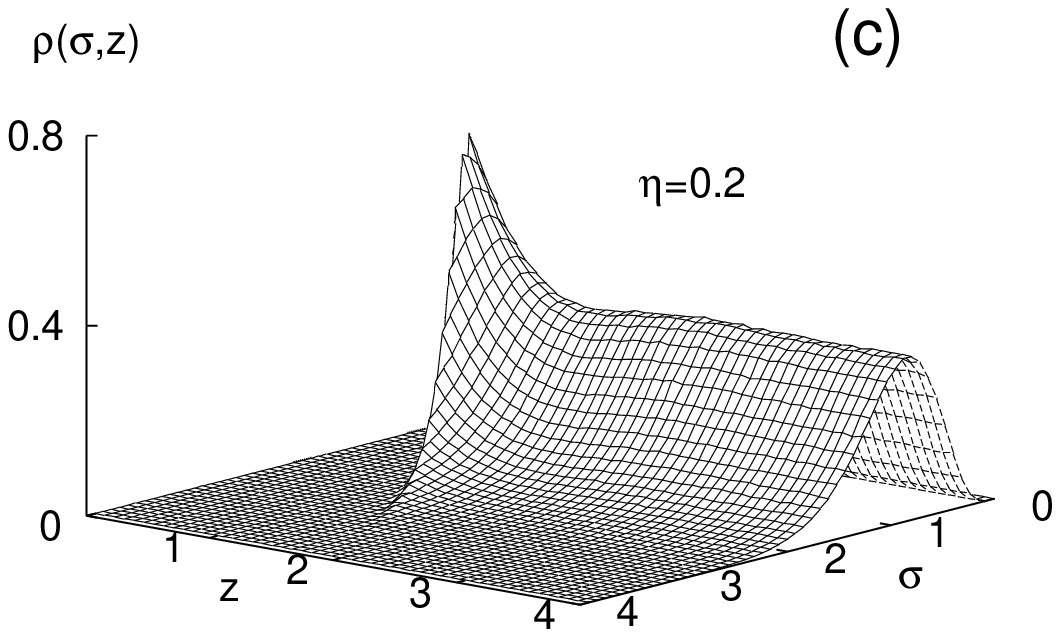,height=5.25cm,width=4.75cm}
\psfig{file=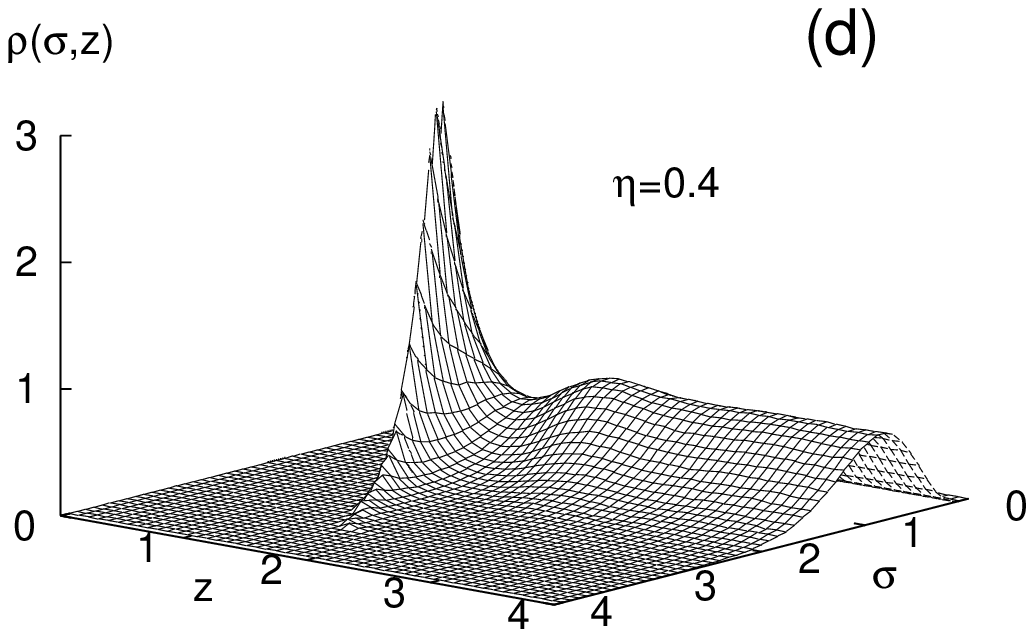,height=5.25cm,width=4.75cm}}
\caption{The particle density distributions $\rho(\sigma,z)$ obtained from the MC simulations. {\bf(a,b)}
Top-hat ($\delta=0.115$), 
{\bf(c,d)} Schulz.}
\label{psz}
\end{figure}

An initial impression of the effects of confinement on the size
distribution can be gained from fig.~\ref{psz}, which shows the form of
$\rho(\sigma,z)$ obtained in the simulations for the narrow top hat and
the Schulz distributions.  Density oscillations normal to a solid
interface are a common feature of confined fluids, and generally
increase in magnitude with increasing volume fraction. In the
polydisperse case, however, fig.~\ref{psz} shows that the character of
the oscillations can depend both on the particular choice of parent
distribution and (for a given parent) on the value of $\sigma$.

\begin{figure}[h]
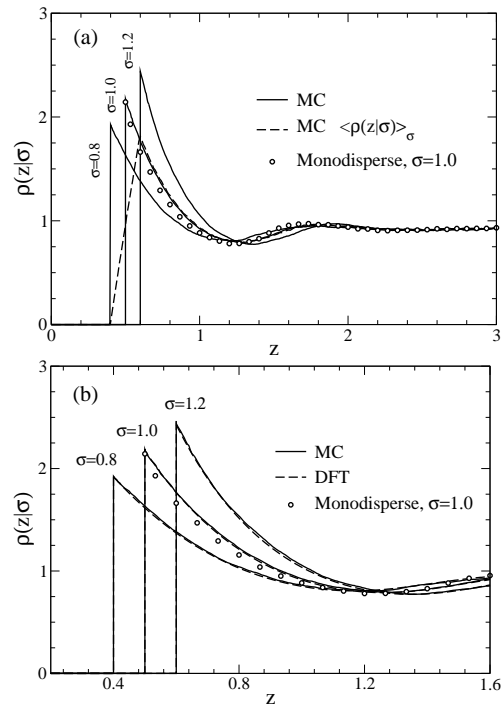

\includegraphics[width=6.5cm,clip=true]{Figs/th02_rhozs_A.eps}
\includegraphics[width=6.5cm,clip=true]{Figs/th02_rhozs_B.eps}
\caption{ {\bf (a)} The spatial distribution $\rho(z|\sigma)$ for
selected values of $\sigma$ for the top-hat distribution ($\delta=0.115$) at $\eta_{b}=0.2$.
{\bf (b)} Detail showing the comparison between MC and DFT results in the region close to the wall. }
\label{fig:th_rhozs_0.2}
\end{figure}

\begin{figure}[h]
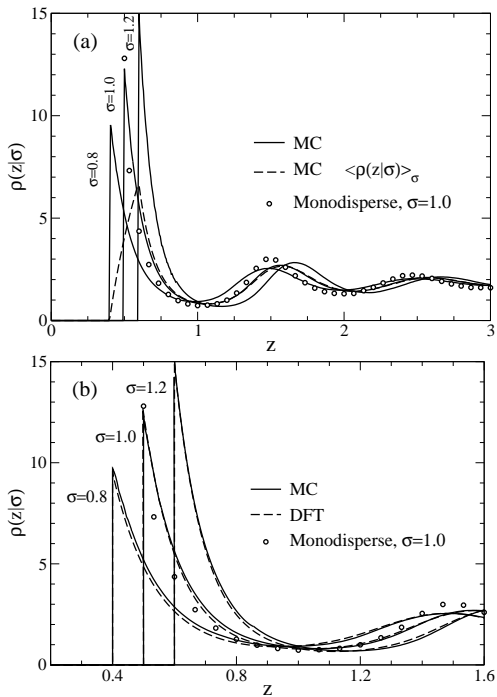

\includegraphics[width=6.5cm,clip=true]{Figs/th04_rhozs_A.eps}
\includegraphics[width=6.5cm,clip=true]{Figs/th04_rhozs_B.eps}
\caption{ {\bf (a)} The spatial distribution $\rho(z|\sigma)$ for
selected values of the particle size for the top-hat distribution
($\delta=0.115$) at $\eta_{b}=0.4$. {\bf (b)} Detail showing the comparison between MC and DFT
results in the region close to the wall.}
\label{fig:th_rhozs_0.4}
\end{figure}

\begin{figure}[h]
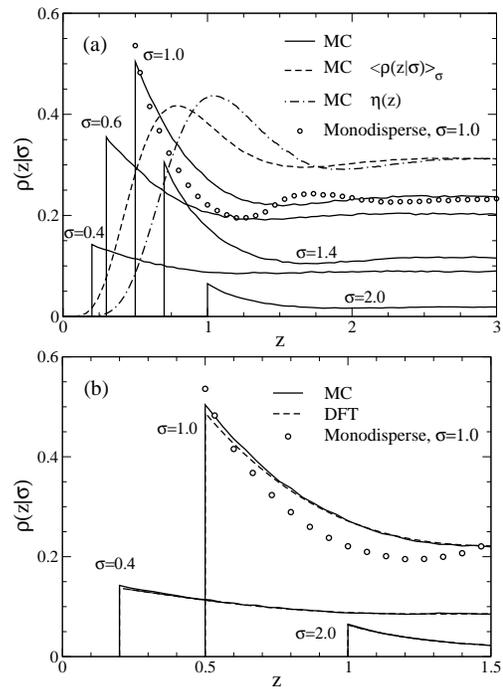

\includegraphics[width=6.5cm,clip=true]{Figs/sc02_rhozs_A.eps}
\includegraphics[width=6.5cm,clip=true]{Figs/sc02_rhozs_B.eps}
\caption{ {\bf (a)} The spatial distribution $\rho(z|\sigma)$ for selected values of the particle size for the Schulz distribution
at $\eta_{b}=0.2$. {\bf (b)} Detail showing the comparison between MC and DFT
results  in the region close to the wall.}
\label{fig:sc_rhozs_0.2}
\end{figure}

\begin{figure}[h]
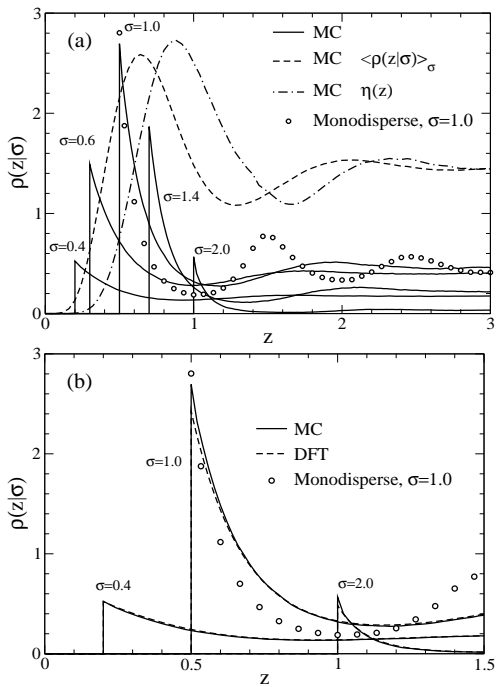

\includegraphics[width=6.5cm,height=4.5cm,clip=true]{Figs/sc04_rhozs_A.eps}
\includegraphics[width=6.5cm,height=4.5cm,clip=true]{Figs/sc04_rhozs_B.eps}
\caption{ {\bf (a)} The spatial distribution $\rho(z|\sigma)$ for selected values of the particle size for the Schulz distribution
at $\eta_{b}=0.4$. {\bf (b)} Detail showing the comparison between MC and DFT results in the region close to the wall.}
\label{fig:sc_rhozs_0.4}
\end{figure}

In order to analyse the data contained in fig.~\ref{psz} in greater
detail, it is instructive to examine orthogonal slices $\rho(z|\sigma)$
and $\rho(\sigma|z)$. Beginning with the former case, examples of
$\rho(z)$ for a selection of values of the particle diameter $\sigma$
are shown for the narrow top-hat distribution at $\eta_{b}=0.2$ and
$\eta_{b}=0.4$ in figs.~\ref{fig:th_rhozs_0.2} and
\ref{fig:th_rhozs_0.4} respectively. The corresponding data for the
Schulz distribution is shown in figs.~\ref{fig:sc_rhozs_0.2} and
\ref{fig:sc_rhozs_0.4}. Included also in these figures are the
predictions of the DFT theory (cf. sec.~\ref{sec:dft}). In all cases
one observes the presence of density oscillations which start from the
bulk density value far from the wall and grow with decreasing $z$ down
to the point $z=\sigma/2$, where the curves terminate abruptly due to
the excluded volume constraint of the hard wall. Also shown for
comparison in each case is the form of $\rho(z)$ for monodisperse hard
spheres (with $\sigma=1.0$) at the same respective volume fraction and
normalized so that its value at $z=L/2$ matches the density profile of
the $\sigma=1.0$ species in the polydisperse system. Additionally,  we
provide the $\sigma$-averaged particle number density
$\langle\rho(z|\sigma)\rangle_\sigma$ as well as (in cases where it
deviates appreciable from the latter), the local volume fraction
$\eta(z)=\frac{\pi}{6}\int d\sigma \rho(z|\sigma) \sigma^3$.

With regard to the general features of the profiles shown in
figs,~\ref{fig:th_rhozs_0.2}--\ref{fig:sc_rhozs_0.4}, we note that an
aspect peculiar to the narrow top-hat parent distribution is that
$\rho(z)$ oscillates with a period $\sigma\simeq 1.0$ (i.e. close to
$\bar\sigma$) for all particle sizes. This contrasts with the case of
the much wider Schulz distribution, where the oscillations do not
appear to be controlled by the average particle size, having a
considerably larger wavelength in $z$. We note further that in all
cases studied, the oscillations in $\rho(z)$ induced by the wall are
more pronounced for the monodisperse fluid than for the polydisperse
ones. Thus the effect of polydispersity appears to be to dampen the
profile oscillations. This is presumably a result of structural
disordering effects deriving from the ability of small particles to
occupy the gaps between large ones. These in turn serve to disrupt the
excluded volume packing effects which underly the density oscillations
commonly observed in monodisperse systems near a hard wall. Within this
qualitative picture, one would thus expect a greater dampening of the
oscillations for the Schulz distribution than for the narrow top-hat,
on account of the disparity in their values of $\delta$. This indeed
appears to be the case, with the top-hat profiles resembling more
closely the monodisperse case.

Turning next to the comparison of the simulation measurements with the
results of the DFT calculations, it is gratifying to note that the
level of agreement is generally excellent, particularly at the lower
volume fraction. Only for the case $\eta_{b}=0.4$ and points close to the
wall do we observe relatively minor systematic deviations, namely a
slightly higher local density for the MC results. 

%We shall postpone a
%fuller discussion on the comparison to sec.~\ref{sec:concs}.

\begin{figure}[h]
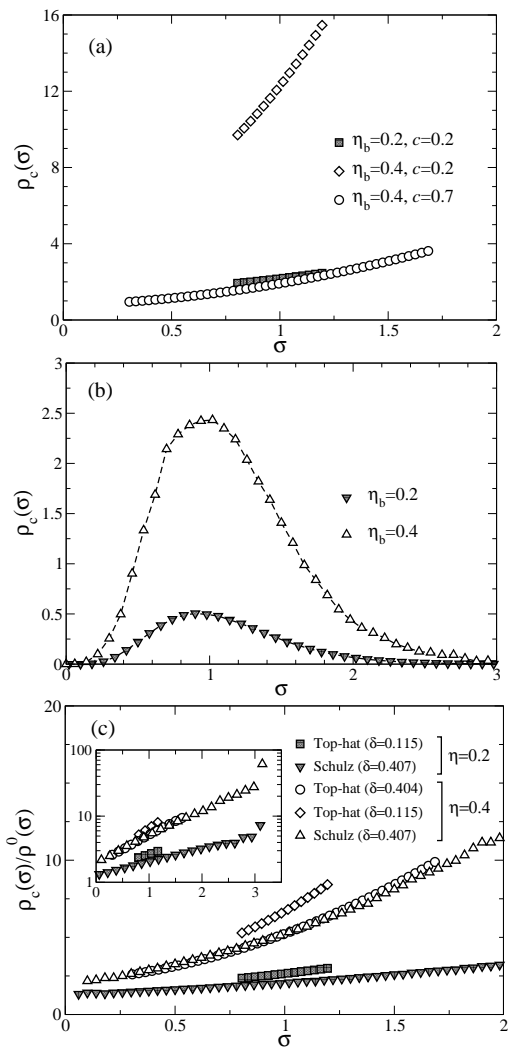

\includegraphics[width=6.5cm,clip=true]{Figs/th_contval.eps}
\includegraphics[width=6.5cm,clip=true]{Figs/sc_contval_bis.eps}
\includegraphics[width=6.7cm,clip=true]{Figs/norm_contval.eps}

\caption{Contact values $\rho_{c}(\sigma)=\rho(\sigma,z=\sigma/2)$. 
{\bf(a)} Top-hat distributions ($\delta=0.115$ and $\delta=0.404$) at
volume fractions $\eta_b=0.2, 0.4$. {\bf(b)} Schulz distribution
$(\delta=0.407)$. {\bf (c)} The contact values relative to the bulk
density. In all cases, lines are merely guides to the eye. Error bars
are smaller than the symbol sizes.}

\label{fig:contval}
\end{figure}

As noted above, the profiles $\rho(z|\sigma)$ are truncated at
$z=\frac{\sigma}{2}$ reflecting the geometrical constraint associated
with the impenetrable wall at $z=0$. The density at this point,
$\rho_c(\sigma)\equiv \rho(z=\frac{\sigma}{2}|\sigma)$ is the so-called
contact value.  In  fig.~\ref{fig:contval} we plot $\rho_c(\sigma)$ for
selected $\sigma$ for both the narrow and wide top-hat, as well as the
Schulz distribution. From the figure, it is apparent that for both
top-hat distributions,  $\rho_c(\sigma)$ increases as a function of
$\sigma$, i.e. the largest particles are preferentially favored over
small ones at the wall. A similar effect occurs for the Schulz
distribution, fig.~\ref{fig:contval}(b), although in this case since
the parent distribution is not constant in $\sigma$, one must form 
$\rho(z=\sigma/2|\sigma)/\rho^0(\sigma)$ in order to expose the density
enhancement at contact (see fig.~\ref{fig:contval}(c)).

The contact values are related to the partial pressures and hence the
total pressure via the sum rule \cite{EVANS89}:

\begin{equation}
\beta p=\int d\sigma \rho_c(\sigma)
\label{eq:pres}
\end{equation}

The resulting estimates of the pressure are presented in
table~\ref{tab:pressure} where they are compared with the predictions
of the moment based expressions of Mansoori {\em et al} (BMCSL)
\cite{BOUBLIK,MANSOORI} and Salacuse and Stell \cite{SALACUSE}. The
former is based on a generalization to mixtures of the
Carnahan-Starling equation of state, while the latter is a
generalization of the Percus-Yevick (PY) theory. The simulation
estimates show good agreement with the analytical equations of state at
the lower volume fraction $\eta_b=0.2$, though at the higher value
there are discrepancies, with the BMCSL equation apparently predicting
the observed pressure more accurately for the top-hat distribution and
the PY equation faring better for the Schulz distribution. The latter
result is in accord with previous simulation studies of the bulk
equation of state for Schulz distributed hard spheres \cite{WILDING02}.
As regards the estimates for the pressure derived from the DFT contact
values, we note the close agreement of these with the PY equation of
state. This finding is, however, not too surprising given that the
specific Rosenfeld functional we employ is tuned to recover the PY
equation in the bulk (see Sec.~\ref{sec:dft}). 

%%vspace*{1truecm}

\begin{table*}
\begin{tabular}{|c|cccc|}  \hline \hline
\makebox[1.5cm]{Parent} &
\makebox[1.5cm]{MC} &
\makebox[1.5cm]{DFT} &
\makebox[1.5cm]{BMCSL} &
\makebox[1.5cm]{PY} \\
\hline
top-hat ($c=0.2, \eta_b=0.2$)  & 0.8718$\pm$0.008 & 0.8748 & 0.8774 & 0.8774 \\
top-hat ($c=0.2, \eta_b=0.4$)  & 4.955$\pm$0.074& 5.074 & 4.971 & 5.180 \\
top-hat ($c=0.7, \eta_b=0.4$)  & 2.809$\pm$0.085 & & 2.86 & 2.966\\
Schulz ($\eta_b=0.2$)          & 0.518$\pm$0.009 & & 0.5149 & 0.5175\\
                               & & 0.516 & (0.5147) & (0.5173) \\
Schulz ($\eta_b=0.4$)          & 2.767$\pm$0.080 & & 2.668 & 2.763 \\
                               & & 2.732& (2.593) & (2.762) \\
\hline \hline
\end{tabular}
\caption{Estimates of the pressure $\beta p$, as calculated from the contact
values (eq.~\ref{eq:pres}) for the MC and DFT calculations for the
top-hat and Schulz parents at the given volume fraction $\eta_b$.
Also  shown for comparison are the predictions of the BMCSL
\cite{BOUBLIK} and PY \cite{SALACUSE} equations of state. The values in
brackets refer to the BMCSL and PY predictions for a Schulz parent
having a lower cutoff in $\sigma$ as employed in the DFT calculations
(see sec.~\ref{sec:dft}).}
\label{tab:pressure}
\end{table*}

The enhancement (relative to the bulk) of the larger particles at
contact (see fig.~\ref{fig:contval}(c)), is attributable to attractive
depletion forces. It is interesting to note that for a given $\eta_b$,
$\rho(z=\sigma/2|\sigma)/\rho^0(\sigma)$ is apparently very similar for
the Schulz distribution and the wider of the two top-hat parents. 
Since both these distributions have a very similar value of $\delta$,
this suggests that the depletion interactions at the wall for a given
$\sigma$ are primarily controlled by the values of $\eta_b$  and 
$\delta$, and are much less sensitive to the overall distribution
shape.

\begin{figure}[h]
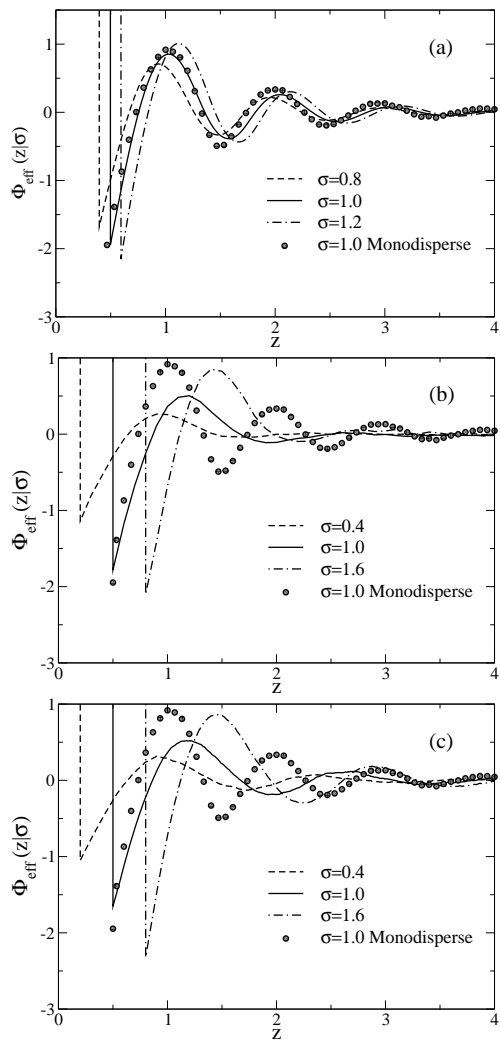

\includegraphics[width=6.5cm,clip=true]{Figs/th_veff.eps}
\includegraphics[width=6.5cm,clip=true]{Figs/sc_veff.eps}
\includegraphics[width=6.5cm,clip=true]{Figs/th07_veff.eps}
\caption{ The effective wall-particle potential at volume fraction
$\eta_{b}=0.4$. {\bf(a)} Narrow top-hat $(\delta=0.115)$;
{\bf(b)} Schulz, $(\delta=0.407)$ and {\bf(c)} Wide top-hat $(\delta=0.404)$
distribution.
}
\label{fig:potential}
\end{figure}

The wall-fluid depletion interaction can be further quantified  (see
for example ref.~\cite{GOETZELMANN99}) in terms of an effective
potential (potential of mean force) via:

\begin{equation} \frac{\rho(z|\sigma)}{\rho^0(\sigma)}=\exp
\left(-\beta\Phi_{eff}(z|\sigma) \right).  
\label{eq:phieff} 
\end{equation}   
Here $\Phi_{eff}(z|\sigma)$ represents the excess
grand potential of a particle of size $\sigma$ at distance $z$ from the
wall with respect to its value in the bulk. Plots of
$\Phi_{eff}(z|\sigma)$ (fig.~\ref{fig:potential}) indeed show the
expected attractive nature of the effective potential at contact, and
the deepening of the well depth with increasing $\sigma$. Note,
however, that the effective potential $\Phi_{eff}$ defined in
eq.(\ref{eq:phieff}) should not be confused with a true two-body
depletion potential. A relationship similar to eq.(\ref{eq:phieff}) has
been used to quantify fluid-wall depletion effects in asymmetrical
binary mixtures comprising large (colloid) particles in a solvent of
small particles (such as polymer). There one defines a depletion
potential between large particles and the wall in the limit in which
the concentration of the large particles \cite{EVANS01,GOETZELMANN99}
vanishes, but at fixed concentration of the smaller ones. By contrast,
in the present case, the depletion forces are not mediated by a
separate solvent but arise from the effects of the distribution of
particle sizes \cite{NOTE1}. Nevertheless similarities are evident with
the form of the true two-body depletion potential observed in binary
mixtures \cite{EVANS01,GOETZELMANN99}. We note further, that the
potentials for the Schulz distribution ($\delta=0.407$) are numerically
very similar to those for the wide top hat ($\delta=0.404$) for
distances $z\lesssim 2.0$. This suggests that our observation (see
above) concerning the apparent insensitivity of the contact
enhancements to the shape of $f(\sigma)$, extends to the form of the
effective potential at significant distances away from the wall. 

\begin{figure}[h]
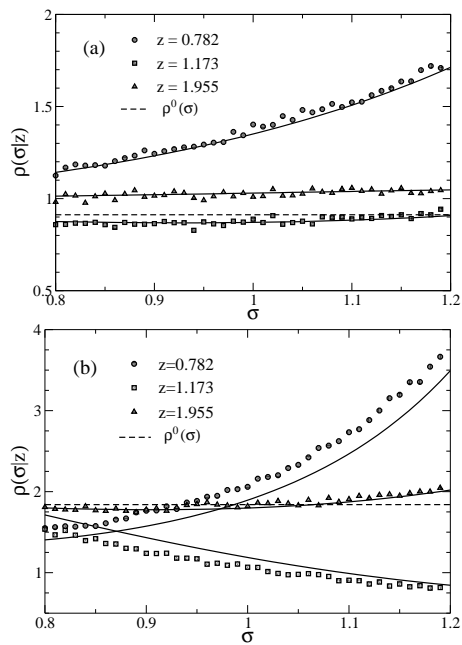

\includegraphics[width=6.0cm,clip=true]{Figs/th02_rhosz.eps}
\includegraphics[width=6.0cm,clip=true]{Figs/th04_rhosz.eps}
\caption{ The size distribution at selected distances $z$ from the wall for the
narrow top-hat distribution ($\delta=0.115$).
{\bf(a)} $\eta_{b}=0.2$, {\bf(b)} $\eta_{b}=0.4$. Symbols: Monte Carlo results. Solid lines: DFT results.}
\label{fig:th_rhosz}
\end{figure}

\begin{figure}[h]
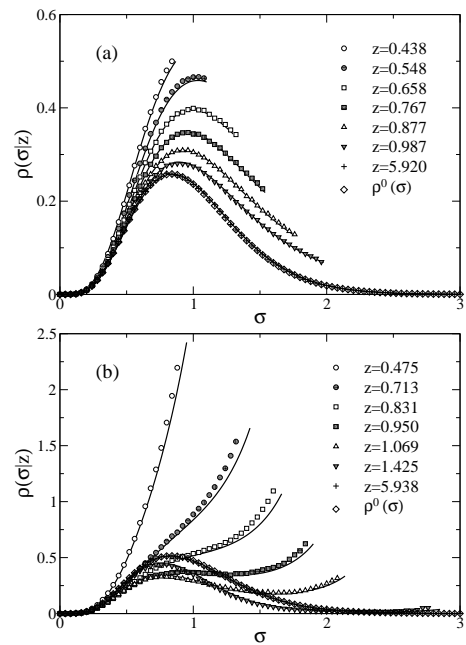

\includegraphics[width=6.0cm,clip=true]{Figs/sc02_rhosz.eps}
\includegraphics[width=6.0cm,clip=true]{Figs/sc04_rhosz.eps}
\caption{ The size distribution $\rho(\sigma|z)$ at selected distances $z$ from the wall for the Schulz
distribution ($\delta=0.407$). {\bf(a)} $\eta_{b}=0.2$, {\bf(b)} $\eta_{b}=0.4$. Symbols: Monte Carlo results. Solid lines: DFT results.}
\label{fig:sc_rhosz}
\end{figure}

We turn next to an analysis of the distribution of particle sizes as a
function of the distance $z$ from the wall, as expressed through the
form of $\rho(\sigma|z)$. In figs.~\ref{fig:th_rhosz} and
\ref{fig:sc_rhosz}, we plot $\rho(\sigma)$ at selected values of $z$
for both the narrow top-hat and Schulz parent distributions at bulk volume
fractions $\eta_{b}=0.2$ and $\eta_{b}=0.4$. Focusing first on the top-hat
case, one observes that as $z$ is increased (starting from near the
wall), the form of $\rho(\sigma)$ alternates between being
monotonically increasing and monotonically decreasing. This implies
that with increasing $z$ the biggest and smallest particles are
alternately favored, an effect known as local size segregation
\cite{PAGON00} (see also below). We note that at the closest allowed
distance from the wall, large particles are always strongly favored.
For large $z$, $\rho(\sigma)$ of course approaches the flat top-hat
parent form.

In contrast to the case of the narrow top-hat, for the Schulz parent the
density of the largest permitted particles (i.e. those allowed by the
geometrical constraint at a given $z$) always exceeds the bulk density
at that $\sigma$ (cf. fig.~\ref{fig:sc_rhosz}). The degree of
enhancement is considerably stronger for $\eta_{b}=0.4$ than
$\eta_{b}=0.2$. In both cases, for sufficiently small $z$, the
distribution increases monotonically. However, as $z$ becomes larger,
the enhancement in the density of the largest particles gradually
reduces, and a peak (associated with the maximum of $\rho^0(\sigma)$)
starts to appear at $z\simeq 1.0$, signalling the crossover to bulk
behaviour. It is interesting to note that for $\eta_{b}=0.4$, the
distribution for $1.05 < z < \sigma_c$ is very broad and exhibits two
maxima. One of these is related to the maximum of the parent
distribution while the other corresponds to the depletion
interaction-induced enhancement of the density of the largest particles
permitted at that $z$. Again, we find that for both parent forms the
DFT appears to capture semi-quantitatively the behaviour observed in
the simulations with some relatively slight discrepancies to be found
only at $\eta_{b}=0.4$.

\begin{figure}[h]
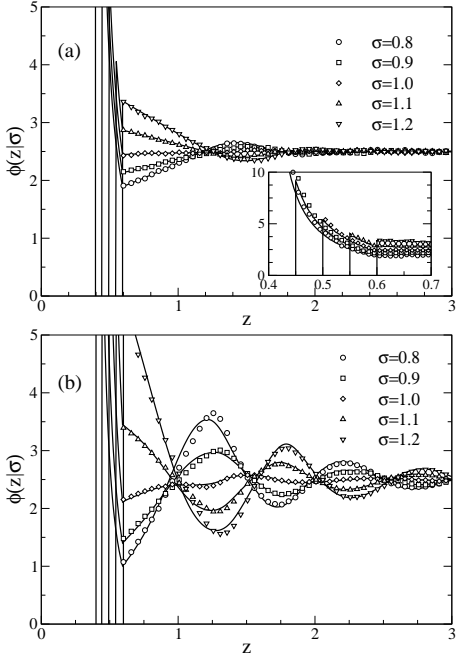

\includegraphics[width=6.0cm,clip=true]{Figs/th02_phi.eps}
\includegraphics[width=6.0cm,clip=true]{Figs/th04_phi.eps}
\caption{The local relative concentration $\phi(z|\sigma)$ for a narrow
top-hat parent distribution ($\delta=0.115$). Data are shown for a
selection of particle sizes $\sigma$. Solid lines are the DFT
predictions. {\bf(a)} $\eta_{b}=0.2$; {\bf(b)} $\eta_{b}=0.4$.}

\label{fig:phi1}
\end{figure}

\begin{figure}[h]
\includegraphics[width=6.0cm,clip=true]{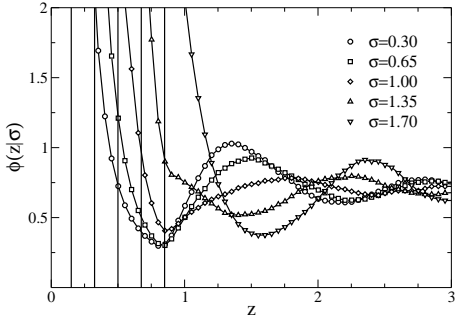}
\caption{The local relative concentration for a wide ($\delta=0.404$) top
hat parent distribution at $\eta_b=0.4$.}
\label{fig:phi_07}
\end{figure}

\begin{figure}[h]
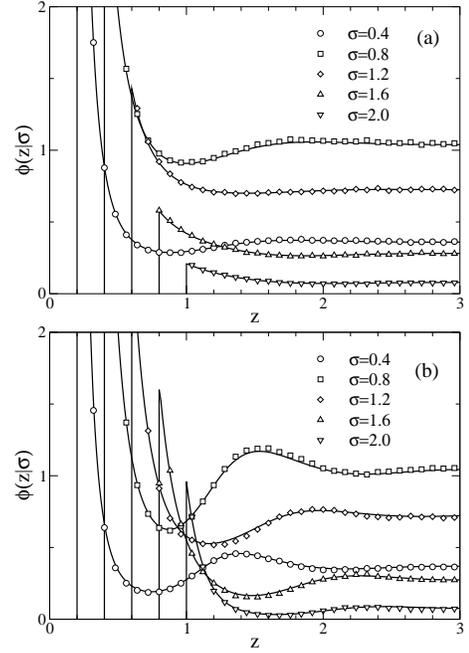

\includegraphics[width=6.0cm,clip=true]{Figs/sc02_phi.eps}
\includegraphics[width=6.0cm,clip=true]{Figs/sc04_phi.eps}

\caption{The local relative concentration $\phi(z|\sigma)$ for the
Schulz parent ($\delta=0.407$). Solid lines are the DFT predictions.
{\bf(a)} $\eta_{b}=0.2$; {\bf(b)} $\eta_{b}=0.4$.}

\label{fig:phi2}
\end{figure}

In order to analyze further the local size segregation effects
identified above, we have considered the spatial dependence of the
local relative concentration defined as: 

\begin{equation}
\phi(z,\sigma)=\frac{\rho(z,\sigma)}{\int_{0}^{\infty}d\sigma'~\rho(z,\sigma')}.
\end{equation} 
This quantity measures the concentration of species of size $\sigma$
at a given $z$. Size segregation is signalled by the appearance of an
oscillatory structure in $\phi(z|\sigma)$. We have obtained the form of
$\phi(z,\sigma)$ for both top-hat distributions
(figs.~\ref{fig:phi1},\ref{fig:phi_07}) and the Schulz distribution
(figs.~\ref{fig:phi2}). The oscillations are more pronounced at the
larger $\eta_{b}$.  For the narrow top-hat, a phase difference of
approximately $\pi$ radians is observed between the largest and
smallest permitted particles and the period is close to the average
particle size \cite{NOTE2}. However, for the wider top hat, the degree
of anti-correlation is somewhat less and the period is extended. For
the Schulz parent, comparison of the curves for $\phi(z|\sigma)$ is
complicated by the fact that at large $z$ each curve converges to a
limit that is proportional to $\rho^0(\sigma)$. Nevertheless, for a
given $\sigma$, clear oscillations are visible, although they die out
more rapidly than for the top hat and do not exhibit a well-defined
anti-correlation.

%\begin{figure}[h]
%\includegraphics[width=6.0cm,clip=true]{Figs/cfr_phi_same_sigma.eps}
%\caption{Same as fig.\ref{fig:phi_07}, showing the effect of the
%distribution width for a top hat parent
%on the local relative concentration of particles of the same size.
%}
%\label{fig:cfr_phi_same_sigma}
%\end{figure}

Finally in this section we turn to our results for the local
polydispersity $\delta(z)$ defined via a generalization of
eq.~\ref{eq:poly}:

\begin{equation}
\delta(z)=\frac
{\left[ \int d\sigma (\sigma-\bar\sigma)^{2} \rho(\sigma|z) \right] ^{1/2}}
{\int d\sigma \sigma \rho(\sigma|z)}.
\label{eq:polygen}
\end{equation}

\begin{figure}[h]
\includegraphics[width=8.5cm,clip=true]{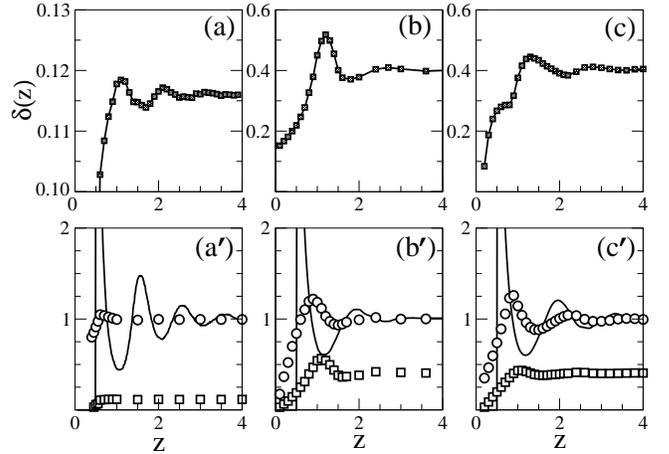}
\caption{The spatial dependence of the local degree of polydispersity
$\delta(z)$ at volume fraction $\eta_{b}=0.4$. {\bf(a)} Top-hat,
$\delta=0.115$; {\bf(b)} Schulz $\delta=0.407$; {\bf(c)} top-hat,
$\delta=0.404$. Also shown for the respective parents ({\bf(a$^\prime$)} etc) are the
mean (circles) and the standard deviation (squares) of the local size
distribution $\rho(\sigma|z)$. Lines show the profile of the mean
species $\rho(z|\sigma=1)$, relative to its bulk value.}
\label{fig:polyz}
\end{figure}

Our measurements of this quantity for both top-hat parents and the
Schulz parent are presented in fig.~\ref{fig:polyz}. One observes
oscillations in $\delta(z)$, derived (cf. eq.\ref{eq:polygen}) from
oscillations in both the local mean and the standard deviation of the
density distribution. The oscillations are stronger for the wider
parents than for the narrow one. Further insight into this behaviour can
be gained from a perturbative analysis of $\delta(z)$ in the limit of
a narrow parent.  We assume that the species deviate slightly from the 
mean species, so that $\sigma=\bar{\sigma} (1+\epsilon)$ with
$\epsilon$ a  small parameter. In this limit, the density profile can
be expressed \cite{PAGON00} as

\begin{eqnarray}
\rho(\sigma,z)&=&\frac{\rho(\bar{\sigma},z)\rho^0(\sigma)}{\rho(\bar{\sigma},\infty)}e^{-\beta[V(\sigma,z)-V(\bar{\sigma},z)]}\nonumber\\
&\times& \left\{ 1+\epsilon[\tilde{c}^\prime(z)-\tilde{c}^\prime(\infty)] \right\}
\end{eqnarray}
where $\rho(\bar\sigma,z)$ is the density profile for the mean species,
while  $\tilde{c}^\prime$ corresponds to the reversible work associated
with slightly changing the size of one particle, a distance $z$ from the wall,
within an otherwise monodisperse systems at the corresponding volume
fraction (see \cite{PAGON00} for further details).  Introducing
this expansion into eq.(16) yields

\begin{eqnarray}
\label{eq:deltanarrow}
\delta(z) &\simeq &\bar{\sigma}\sqrt{\frac{\rho(\bar{\sigma},\infty) \overline{\epsilon^2}}{\rho(\bar{\sigma},z) n_0}}\nonumber\\
         &\times&\left\{  1+\left[\frac{\overline{\epsilon^3}}{2\overline{\epsilon^2}}-\overline{\epsilon^2}\right][\tilde{c}^\prime(z)-\tilde{c}^\prime(\infty)]\right\}\;,
\end{eqnarray}
at distances where the wall potential is negligible.

Eq.~\ref{eq:deltanarrow} indicates that for a narrow parent the local 
deviations in the width of the distribution depend on second and higher
moments of the parent. As a result, in this limit the local variations
of the mean density make a relevant contribution to $\delta$. This is
confirmed by fig.~\ref{fig:polyz}(a) which shows that the oscillations
are indeed anticorrelated with those of the mean species. In fact, the
amplitude  of the term $\tilde{c}^\prime(z)-\tilde{c}^\prime(\infty)$
is around $0.004$ for a flat parent with $\delta=0.115$ and around
$0.054$  for $\delta=0.404$. Since the deviations of $\tilde{c}^\prime$
are never larger than unity for the parameters considered (see e.g.
Fig. 2 in ref.\cite{PAGON00}), this explains the relative weakness of
the oscillations in fig.~\ref{fig:polyz}(a). It also helps to explain
the behavior of fig.~\ref{fig:polyz}(b) (although in this case one is
far from the narrow limit): for a Schulz distribution
${\overline{\epsilon^3}}=2{\overline {\epsilon^2}}^2$ and hence the
correction in  eq.(\ref{eq:deltanarrow}) vanishes. Accordingly, in a
Schulz distribution the shape corrections to $\delta(z)$ will become
significant only at large $\delta$. In fact, comparing $\delta(z)$ for
$\delta=0.404$ between the Schulz and top hat parents
(fig.~\ref{fig:polyz}), one can see that for a Schulz the shape of the
local degree of polydispersity follows more closely the density profile
of the mean species. 

Although we have used a definition of the local polydispersity that
generalized the conventional form~(eq.\ref{eq:poly}) used to describe bulk
parent forms, it is interesting to note that, at least for 
inhomogeneous systems, a definition of $\delta(z)$ based on the local 
concentration, $\phi(\sigma,{\bf r})$, rather than on the local number
density, serves to better illustrate the role of polydispersity
effects. Accordingly we form a narrow limit expansion of

\begin{equation}
\widehat{\delta}(z)\equiv\frac{\left[\int d\sigma (\sigma-\bar{\sigma})^2 
\phi(\sigma,z)\right]^{1/2}}{\int d\sigma \sigma \phi(\sigma,z)}\:,
\end{equation}
which yields,
\begin{equation}
\widehat{\delta}(z)\simeq 
\bar{\sigma}\sqrt{\overline{\epsilon^2}}
\left\{1+\left[\frac{\overline{\epsilon^3}}{2 
\overline{\epsilon^2}}-\overline{\epsilon^2}\right]\left[\tilde{c}^\prime(z)-\tilde{c}^\prime(\infty)\right]\right\}\:.
\end{equation}
One sees that this last expression does not depend on the local density of
the mean species;  accordingly we would generically expect a much weaker variation
of $\hat{\delta}(z)$  for a Schulz, and for other asymmetric
parents. This alternative definition of the local polydispersity thus
highlights the fact that  in the narrow limit the local segregation of species
tends to preserve the width of the parent imposed in the bulk.

\section{Discussion and conclusions}
\label{sec:concs}

In summary we have employed specialized MC simulations techniques and
DFT to study the effects of a hard wall on the properties of
polydisperse hard spheres. Bulk phase (parent) distributions of the
top-hat and Schulz forms were considered. In the former case, degree of
polydispersity $\delta=0.115$ and $\delta=0.404$ were studied, while in
the latter case, $\delta=0.407$ was used. In all cases the density
distribution $\rho(\sigma,z)$ was obtained and analyzed for bulk volume
fractions $\eta_{b}=0.2$ and $\eta_{b}=0.4$. 

The original motivation for this study was a desire to gauge the
accuracy of the DFT predictions of ref.~\cite{PAGON00} via a
like-for-like comparison with simulation. The results presented in
sec.~\ref{sec:results} demonstrate that the agreement is extremely
good. Indeed, the DFT calculations provide a quantitatively accurate
description of the system properties at $\eta_{b}=0.2$ for both parent
forms studied. Even for the higher volume fraction $\eta_{b}=0.4$, the
agreement is semi-quantitative. Clearly this finding bodes well for the
future utility of DFT in investigating other polydispersity related
issues in inhomogeneous fluid.

Beyond this, the present study extends that of ref.~\cite{PAGON00} by
considering the nature of the attractive depletion interactions in the
vicinity of the hard wall. Analysis of contact values showed a clear
preference for the largest particles to occupy positions in which their
surfaces touch the wall. Indeed for the Schulz distribution, the
contact values of particles at the cutoff exceeded that in the bulk by
over two orders of magnitude. Interestingly, at a given $\eta_b$, the
degree of contact enhancement as a function of $\sigma$, seems quite
insensitive to the distribution shape (cf. fig.~\ref{fig:contval}(c))
for a given $\delta$. This finding was found to extend to the form of
the effective potentials (fig.~\ref{fig:potential}) near the wall.
Clearly however our comparison is far from exhaustive in this regard
and further studies would be required to determine whether this
observation applies more generally.

The interplay of the $\sigma$-dependent attractive depletion forces and
the geometrical constraint imposed by the wall was found to radically
alter the local size distribution. Indeed in the case of the Schulz
parent at $\eta_{b}=0.4$, a minimum in $\rho(\sigma|z)$ was observed to
develop within a certain range of $z$. Another interesting feature,
was the appearance of oscillatory structure in the local degree
of polydispersity $\delta(z)$. This was accompanied, very close to the
wall, by considerable reduction in the local degree of polydispersity
with respect to the bulk. It is tempting to speculate that this latter
finding may have some bearing on recent experimental observations of
wall induced freezing in a polydisperse system of colloidal hard
spheres that apparently forms a glass in the bulk \cite{DULLENS}.
Polydispersity is known to hinder freezing in colloidal systems
\cite{PUSEY}, and consequently any narrowing of $\rho(\sigma)$ due to
the presence of a wall may serve to promote the formation of an ordered
crystal phase.

With regard to the size segregation effects originally reported in
ref.~\cite{PAGON00}, we find that the anti-correlation in the
oscillations of the local concentration between the largest and
smallest permitted particles is significantly more pronounced for small
$\delta$ than for large $\delta$ at a given $\eta_{b}$. Furthermore as
for small $\delta$, the period of the oscillations in $z$ appears to be
close to that of the mean particle diameter. Both these findings are
in accord with the predictions of perturbative calculations
\cite{PAGON00}. For large $\delta$, by contrast, the present work shows
that distribution shape effects {\em are} important in determining the
character of size segregation effects. Thus for the Schulz parent at
$\delta\approx 0.4$, the degree of anti-correlation is much less than
for the corresponding top-hat parent. Additionally, in the latter case
the period of the oscillations appear to be more consistent with the
average particle size than for the Schulz distribution. This suggests
that it is the presence in the Schulz parent of relatively large higher
order moments which leads to this increased disruption of
size-segregation effects. Analogous observations apply to the density
profiles $\rho(z|\sigma)$ themselves. In all cases, (but especially for
the Schulz parent), the presence of polydispersity was found to dampen
the density oscillations significantly compared to the monodisperse
profile $\rho(z)$.

%The origin of the concentration oscillations is presumably traceable to
%the enrichment (relative to the bulk) of large particles near the wall,
%thereby displacing smaller ones to the next layer, and the knock-on
%correlations between particles of different sizes in successive layers.

Turning finally to the outlook for further related work, an interesting
open question is whether size segregation effects analogous to those
considered here for an inhomogeneous system might also occur in bulk
polydisperse fluids i.e. whether pairs of hard spheres of prescribed
sizes $\sigma$ and $\sigma^\prime$ are preferentially found with
certain separations. This could be investigated by appeal to the
measured form of the pair distribution function
$g(r,\sigma,\sigma^\prime)$. We hope to report on this matter in a
future publication.

\vspace*{5mm} 
\acknowledgments

This work was supported by the EPSRC, grant number GR/S59208/01. I.P.
acknowledges financial support from DGICYT of the Spanish Government. 
NBW acknowledges a useful discussion with R. Evans.


\begin{thebibliography}{99}
\bibitem{NOTE0} The term ``parent'' is chosen to emphasize that under
conditions of confinement, the local distribution may differ from the bulk
form from which it derives. Similar effects occur when a polydisperse
fluid phase separates into coexisting "daughter" phases whose local
distributions differ from that of the parent \cite{SOLLICH02}.
\bibitem{NOTE2}  It is worth stressing that these phase shifts are not
simply a consequence of the fact that the centers of smaller particles
can approach the wall more closely than larger ones. They remain even
if one represents the distance from the wall in terms of a shifted
coordinate $z^{*}(\sigma)=z-\sigma/2$.
\end{thebibliography}
\end{document}